\documentclass[useAMS,usenatbib]{mn2e}
\usepackage{graphicx}

\title[Spectropolarimetery of umbral fine structures]
{Spectropolarimetery of umbral fine structures from Hinode: Evidence for magnetoconvection}
\author[Bharti at al.]{Lokesh Bharti$^{1}$\thanks{E-mail:
lokesh\_bharti@yahoo.co.in}
\thanks{Currently at Max-Planck Institute for Solar System Research, 37191 Katlenburg-Lindau, Germany}, Chandan
Joshi$^{1}$, S.N.A. Jaaffrey$^{1}$ and Rajmal Jain$^{2}$\\
1. Department of Physics, University College of Science, Mohanlal Sukhadia University,
Udaipur, 313001, India\\
2. Physical Research Laboratory, (Department of Space, Government of India) Navrangpura, Ahmedabad 380 009, India\\
}
\begin{document}

\maketitle

\label{fristpage}

\begin{abstract}
We present spectropolarimetric analysis of  umbral dots and a light bridge fragment
 that show dark lanes in G-band images. Umbral dots show
upflow as well as associated positive Stokes {\it V} area asymmetry in their central parts.
Larger umbral dots show
down flow patches in their surrounding parts that are associated with negative Stokes {\it V} area asymmetry.
 Umbral dots show weaker magnetic field in central part and higher magnetic field
in peripheral area. Umbral fine structures are much better visible in total
circularly
polarized light than in continuum intensity. Umbral dots show a temperature deficit above dark
lanes. The magnetic field
inclination show a cusp structure above umbral dots and a light bridge fragment. We compare our
observational findings with 3D magnetohydrodynamic
simulations.
\end{abstract}


   \maketitle

\section{Introduction}
Two models of the umbral dots (UDs) are under discussion these days.
The first is the cluster model (Parker, 1976 and Choudhuri, 1986), that suggests that UDs
are the top of the intrusion of field free material between the flux tubes beneath the sunspot. The second model is
known as the monolithic model (Weiss, 2002 and reference therein)
and suggest that UDs show up because of magnetoconvection in monolithic flux tube.
Recent simulations by Sch\"ussler and V\"ogler (2006) with gray radiative
transfer show UDs appearing due to magnetoconvection in strong background magnetic
field. Knowledge of the nature of UDs is essential to understand the energy
transport from below the sunspot (see reviews from Solanki, 2003 and Thomas \& Weiss, 2004
and reference therein on the subject).

Bharti {\it et al.} (2007a) analyzed Dopplergrams obtained from filtergraph data and they found a
correlation between intensity and velocity in UDs, which suggests a magnetoconvective
origin. Using high quality G-band images from Hinode, Bharti et al.
2007b reported on dark lanes in UDs. These separate observational findings
are compatible with some aspects of simulations by Sch\"ussler and V\"ogler (2006).
Socas-Navarro {\it et al.} (2004) analysed peripheral UDs in detail from spectropolarimetric
data  and find higher temperature ($\sim$ 1 kK), weaker field ($\sim$ 500 G), small upflow ($\sim$ 100 m$^{-1}$)
and more inclined field ($\sim$ 10$^{\circ}$) in UDs.

 In this article we present spectropolarimetric analysis of dark laned
umbral fine structure from Hinode spectropolarimetric data.
The high polarimetric sensitivity and spatial resolution
achieved by Hinode spectropolarimeter now it became possible to compare observational
results directly with predictions of numerical simulations
(Rezaei {\it et al.}, 2007, Sainz Dalda and Bellot Rubio, 2008).

\section{Observations and inversion technique}

We used spectropolarimeteric data obtained by the spectropolarimeter onboard
the Hinode (Kosugi {\it et al.}, 2007) on December 12, 2006. The four Stokes profiles of
the two iron line pairs at 630.15 nm (L\'ande factor {\it g}=1.67) and 630.25 nm ({\it g}=2.5) were recorded
for the active region 10930 close to the disk center ($\mu$=0.99). We used fast map.
The integration time for fast map was 3.2 sec.
The field of view comprises an area of 295$\prime$$\prime$ $\times$ 162$\prime$$\prime$
. The spatial sampling for the fast map
was 0$\prime$$\prime$.316 along the slit and 0$\prime$$\prime$.295 in the scanning direction.
The spatial resolution of the resulting spectropolarimetric map is approximately 0$\prime$$\prime$.6
for the fast map
with the spectral sampling at 2.15 pm. The calibration of the SP data is described
by Ichimoto {\it et al.} (2007). We used the Solar-Soft pipeline to calibrate
the SP data.

To derive accurate photospheric height stratification of the temperature (T), magnetic field strength (B), line of sight
velocity (V$_{los}$), and inclination ($\gamma$), we employ the SIR code (Ruiz Cobo
and Del Toro Iniesta 1992). This code presumes hydrostatic equilibrium and local
thermodynamic equilibrium (LTE).  By solving numerically the radiative transfer (RTE) equation for polarized light the inversion
code SIR computes the synthetic Stokes profiles. The optimal parameters for the model were determined iteratively.
The difference between the observed and synthetic Stokes profiles was minimized using a non-linear, least square Marquardt's
algorithm. The values of the physical parameters are computed at only a few grid points called nodes instead of computing
at all optical depths of the model. For rest of depths, they are
approximately computed by the cubic-spline interpolation between the equidistantly distributed grid points. We perform
the SIR inversion with only one magnetic component, for which we allow 5 nodes in T($\tau$), 3 for B($\tau$), V$_{los}$($\tau$),
 and $\gamma$($\tau$).

G-band time series obtained in the broadband filter  were used to follow the evolution of
the sunspot fine structure as seen in the SP maps (see Bharti {\it et al.}, 2007b). Wiener
filtering was applied to the G-band images for the point spread function correction of
telescope. Understanding of the evolution of the umbral
 fine structure is necessary as they may have common physical origin (Rimmele 2008, Katsukawa et al. 2007).
 Here we would like to mention that it is our aim to investigate dark
lane in UDs as reported by  Bharti {\it et al.} (2007b) and the SP fast scan at
0$\prime$$\prime$.6  spatial resolution cover similar features.

The calibrated Stokes profiles were used to create maps of
total circular polarization (TCP) and Stokes {\it V} area asymmetry (Bellot Rubio {\it et al.}, 2007).

\begin{figure}
\vspace{-138mm}
\hspace{11mm}
  \includegraphics[width=225mm]{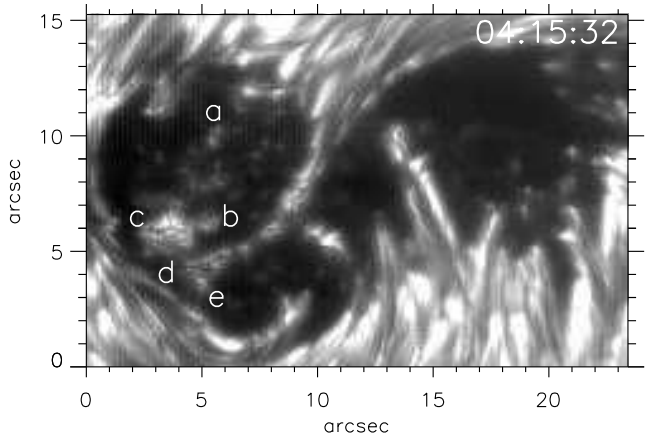}
\vspace{6mm}
\caption{A G-band image as observed in the broad-band filter on SOT at 04:15:32 UT. The image is byte-scaled and shows the fine
structure of the sunspot around the time that our spectropolarimetric maps were taken. }
  \label{models}
\end{figure}

\begin{figure}
\vspace{-138mm}
\hspace{18mm}
  \includegraphics[width=225mm]{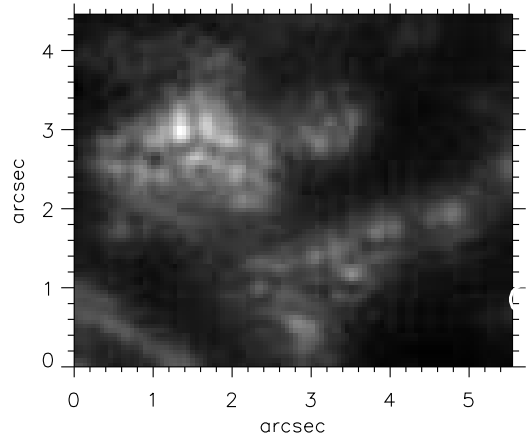}
\vspace{6mm}
\caption{Enlarged view of an area of Figure 1. It clearly shows dark lanes inside the umbral fine structure. }
  \label{models}
\end{figure}

\begin{figure}
\vspace{-260mm}
\hspace{12mm}
 \centering
\includegraphics[width=70mm,angle=0]{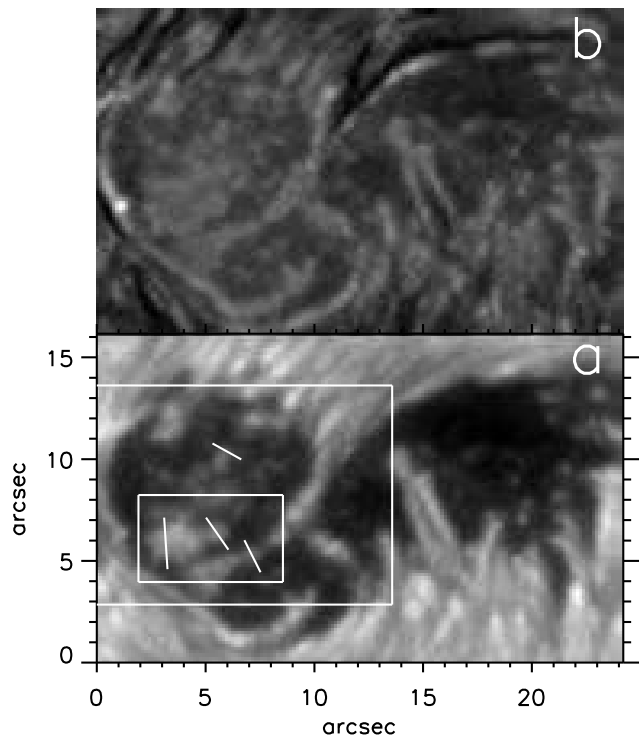}
\vspace{8mm}
\caption{ Maps showing continuum intensity (a) and total circular polarization (b) for the emarging sunspot that rises below a
developed sunspot in AR 10930. Marked in white in panel (a) are the location
where stratifications of various physical parameters along cut were measured. All maps are up scaled two times using cubic spline interpolation.
The outer rectangle shows the field of view that was subjected to the SIR inversion and the inner rectangle shows the
region that was subjected to area asymmetry measurement.}
\end{figure}

\section{Results}

We have chosen G-band images, whose timing was close to the SP scans. Figure 1 shows one of
the G-band images taken close to the fast SP scan time  at 04:15:32 UT that cover emerging sunspot
in the SP map. The G-band time series shows UD `a' has been forms from a peripheral umbral dot (PUD)that fragmented in two UDs.
The UD `b' emerges from a bright bands (Bharti {\it et al.}, 2007b).
The bright band fragments and a UD forms, it grows gradually and shows a threefold dark lane.
At 04:15:32 UT it shows a central bright structure surrounded by a dark ring and five
fragments separated by dark lanes. The time series show that this UD fragments and again
converts in to a bright band. A larger UD `c' forms from a bright band that show complex shapes
during its evolution. In Figure 1 it shows clearly threefold dark lanes.  The light bridge that
develops from the dark cored penumbral filament shows central dark lanes and its
fragments show dark lanes.
At the head of the light bridge a triangular shaped bright structure `d' is seen that
forms from the light bridge fragments in upper part and conglomeration of an UD that
appears from the diffuse background.  This UD is marked by `e'. Individual UDs are seen. However, their
boundary is not clearly visible. Close to the vertex of the triangular structure `d' a light bridge fragment is seen
that shows dark lane. Dark lanes in UDs and light bridge fragments are visible but only UD `c'
shows more clearly dark lane in the G-band image. Figure 2 shows enlarged view of
a part of Figure 1 that shows dark lanes in umbral fine structures very clearly.

Figure 3 illustrates map of the emerging sunspot taken from the SP fast scan.
Panel (a) of Figure 3 shows the continuum intensity map; the fine structure is very similar to that of the G-band image.
Comparision with the G-band image in Figure 1 illustrates that it is reasonable to analyse these
fine structures  using the spectropolarimetric fast scan at 0$\prime$$\prime$.6 resolution.
However, the UD marked by an `a' in Figure 1 shows clear structure in this figure, but not at the lower resolution in Figure 3.
The light bridge fragments are visible in the continuum image of the SP map
(Figure 3(a)). Figure 3(b) depicts TCP map of the same emerging
sunspot. One can see that umbral fine structure such as UDs and diffuse background are much more
prominent in the polarized light than in the continuum intensity. Appearance of umbral fine
structure more prominently in the polarized light suggests that there is no one to one
relationship between the umbral continuum intensity signatures and umbral magnetic fields.
Hence, the umbral inhomogeneities can not be completely characterized only based on the intensity
measurement of umbral fine structures and should be treated cautiously. Similar conclusion was drawn by Bellot Rubio {\it et al.}
(2007) for dark cored penumbral filament from the SP normal map at 0$\prime$$\prime$.3 resolution.
This also helps us to find dark nuclei for zero velocity reference.

\begin{figure*}
 \centering
 \includegraphics[width=130mm,angle=0]{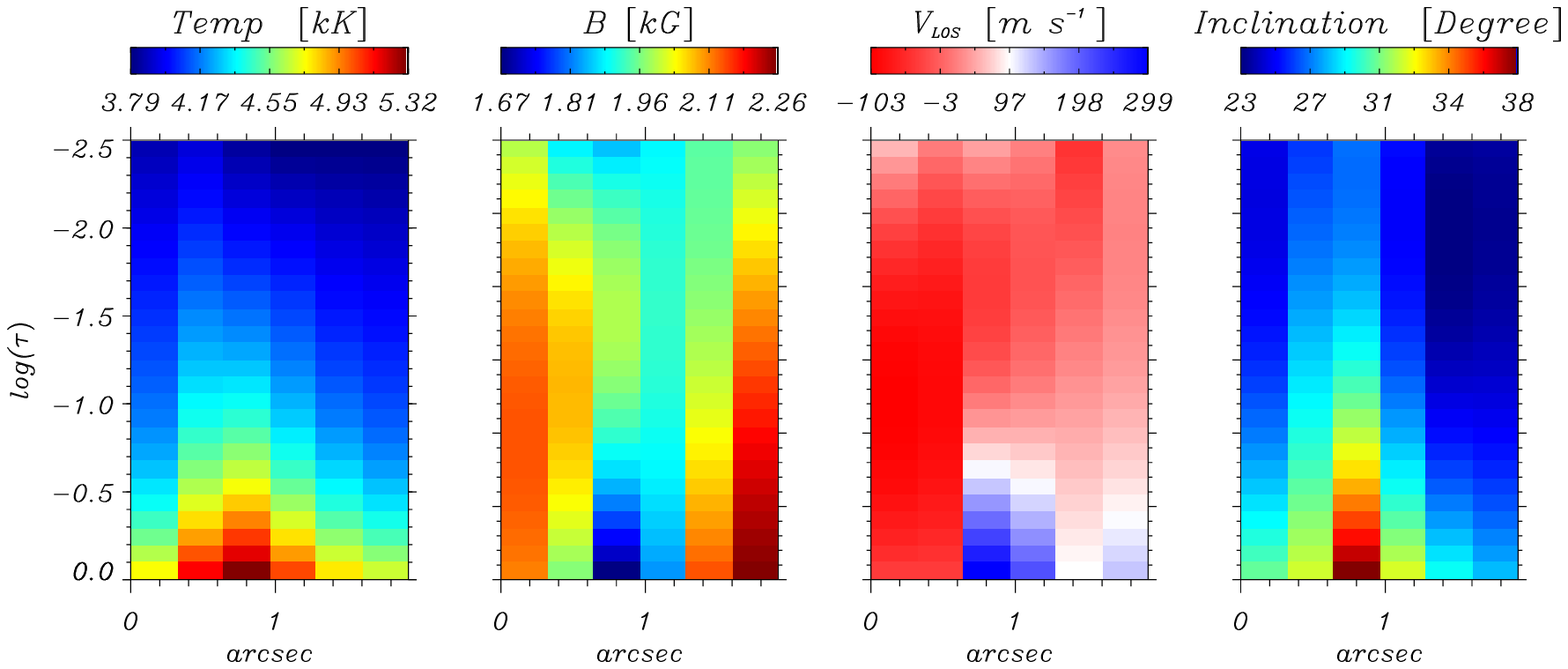}
\vspace{-3mm}
\end{figure*}

\begin{figure*}
 \centering
 \includegraphics[width=130mm,angle=0]{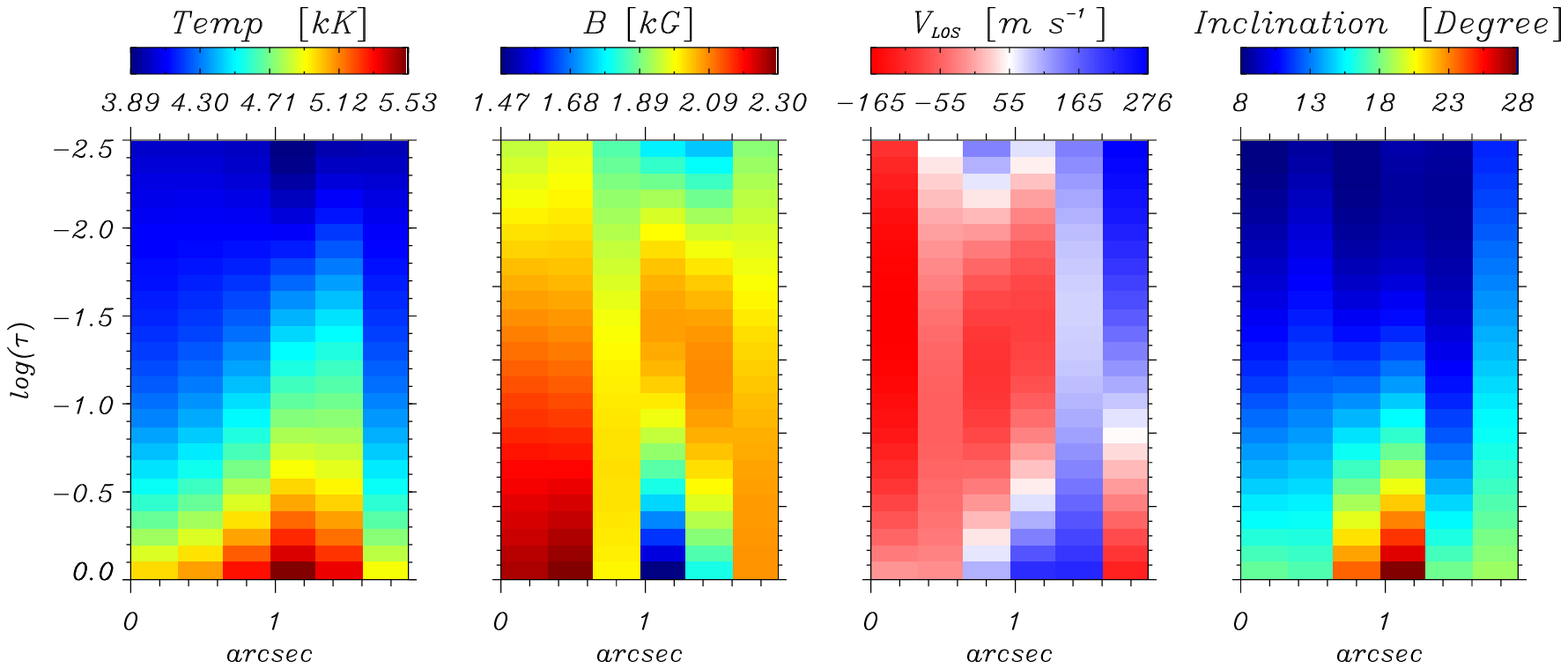}
\vspace{-3mm}
\end{figure*}

\begin{figure*}
 \centering
 \includegraphics[width=130mm,angle=0]{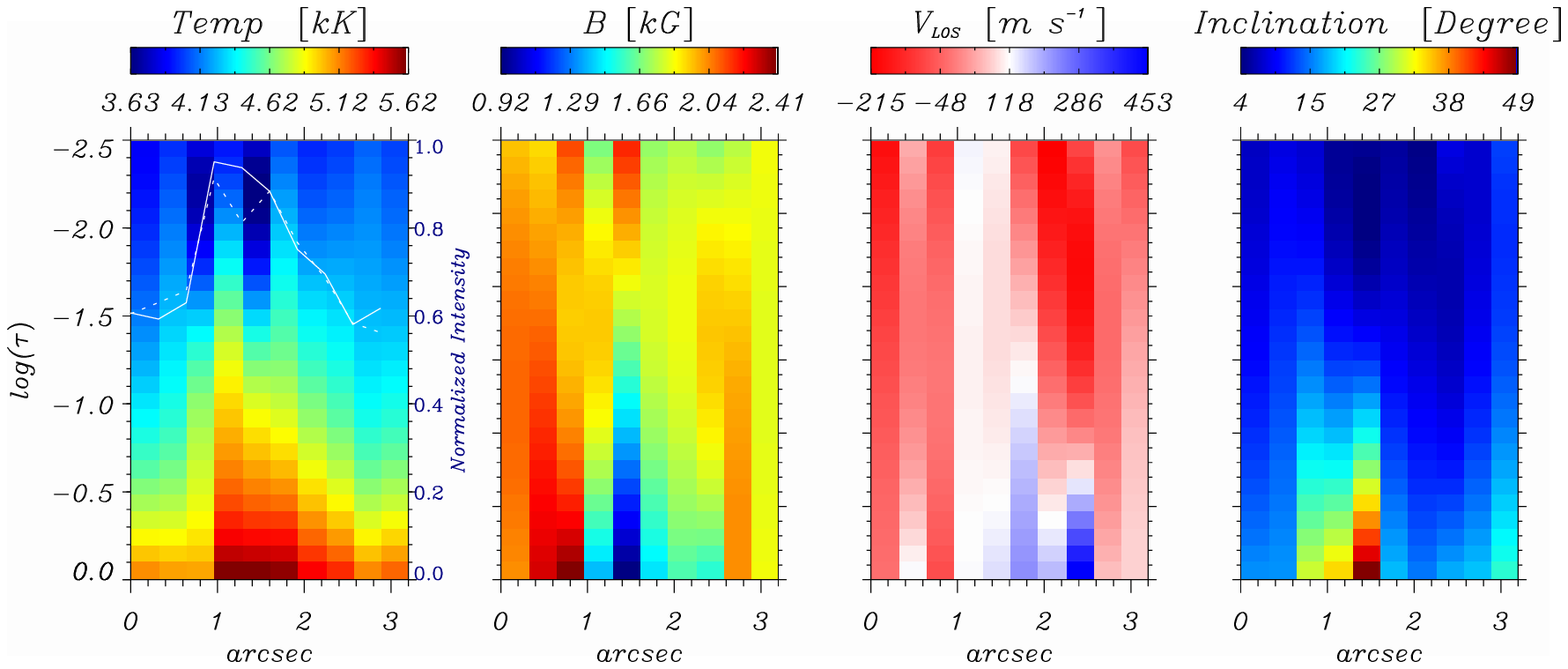}
\vspace{-3mm}
\end{figure*}

\begin{figure*}
 \centering
 \includegraphics[width=130mm,angle=0]{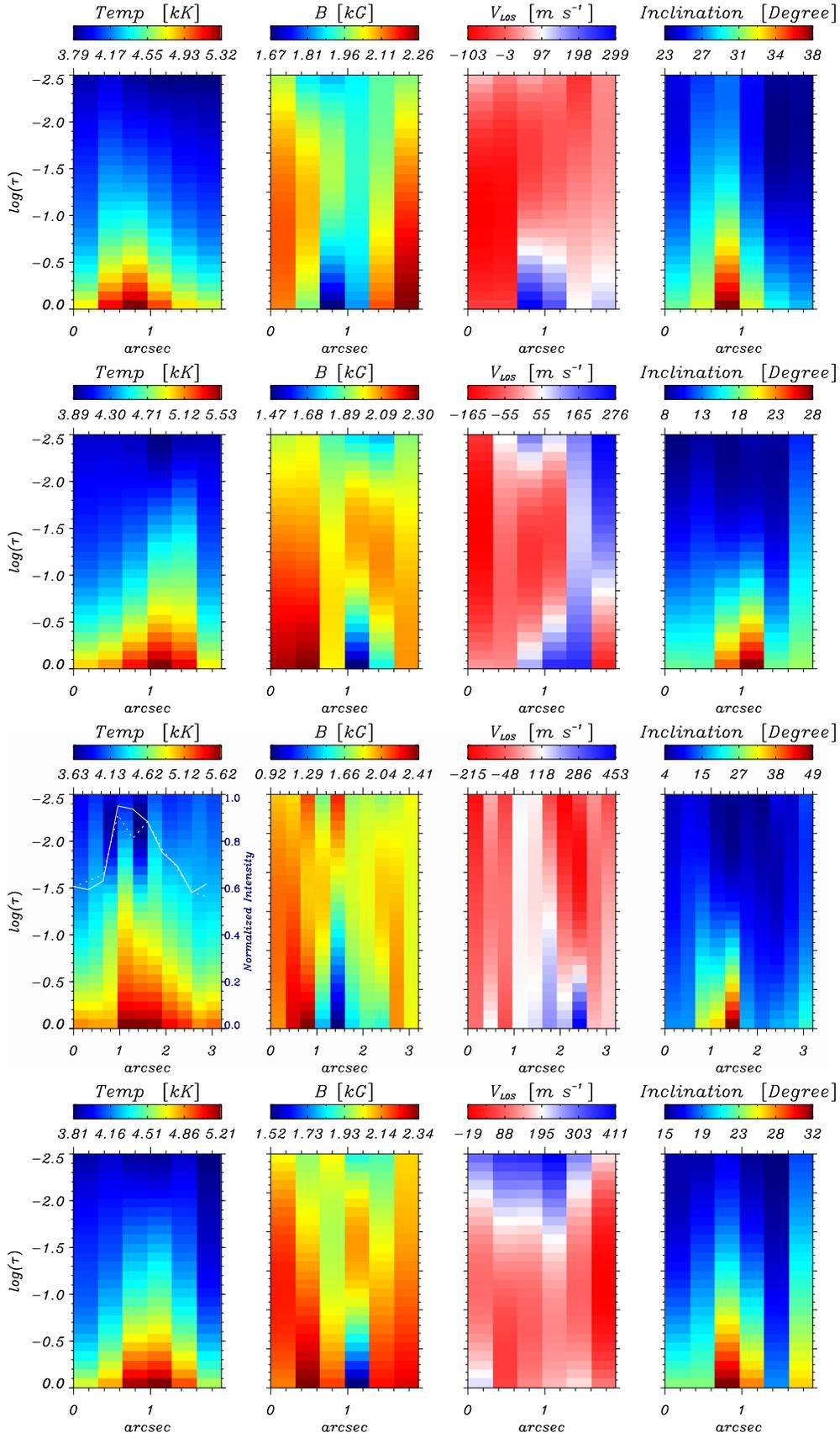}
\vspace{-3mm}
\caption{From left to right: temperature, magnetic field strength, LOS velocity and inclination. From top to bottom: UDs a, b, c and
light bridge.
For UD `c' continuum (solid line) and core intensity (dotted line) overplotted. Positive velocity corresponds to upflow and negative
 velocity to downflow.}
\end{figure*}

      Figure 4 shows the height stratification of plasma parameters according the SIR inversion. This Figure
shows temperature (T($\tau$)), magnetic field strength (B($\tau$)), line-of-sight velocity
 (V$_{los}$($\tau$)) and  inclination ($\gamma$($\tau$)) as a function of altitude and horizontal distance across the lines in panel (a)
 of Figure 3 for umbral fine structures for optical depths 0$>$log($\tau$)$>$-2.5. We find these parameters
reliable for log($\tau$)$<$-2.0; i.e. error are higher in the upper atmosphere. From the top downwards, the plot shows umbral dots
`a', `b', `c' and the light bridge fragment.

The first column of Figure 4 illustrates the temperature stratification of the observed UDs and the light bridge fragment. At $\tau$=1, UDs
 `a', `b', `c' and the light bridge fragment are 780 K, 810 K, 1220 K and 870
 K respectively hotter than the coolest part of the sunspot. This is in agreement with Sobotka and Hanslmeier (2005) who reported from
 two color photometry, on average,
UDs are about 1000 K hotter from the coolest area in the umbra. In all cases these structures are cooler in the higher layers.
In case of UD `c' a temperature drop can be seen around log($\tau$)=$-1.5$, which is colocated with dark lane. This dark lane
has higher contrast in the line core of the 630.25 nm line as shown by the dotted line overplot in Figure 4. However, as shown by the solid line
overplot in Figure 4 this dark lane is not visible in the continuum of the 630.25 nm line. This
indicates that dark lanes have lower temperature compared to the surrounding at the same optical depth.
This is in agreement with Sch\"ussler and V\"ogler (2006) and Spruit and
Scharmer (2006). They suggest that the surfaces of constant optical depths are elevated in these structures, so that they correspond
to lower temperature. This dark lane can be produced by the effect of the density and the gas pressure. The SIR code takes those into
account only approximately under the approximation of hydrostatic equilibrium.
Ruiz Cobo and Bellot Rubio (2008) modeled dark lanes in penumbral filaments and suggested dark lanes are
produced by locally enhanced density and pressure that shift the $\tau$=1 level to higher layers.
All three thermodynamic parameters (temperature, density and gas pressure) are likely to play a role, as suggested by Borrero (2007).
The dark lane is identified clearly in the temperature map at higher layers (not shown here), which consistent with a elevated $\tau$=1 level lies
 above UD in the higher layers.
 We find a temperature variation in the UD `c', which suggests that multifold dark lanes are the manifestation of a temperature
deficit. To our knowledge this is the first observational evidence of temperature deficit in dark lanes of UDs.

      The second column of Figure 4 shows the magnetic field strength stratification of these structures. The field strength in UDs
decreases rapidly with depth. On the other hand, background field strength increases slightly. We find there is a strong
difference between the values in the central part and the peripheral part of UDs. The magnetic field is less in the central part and higher
in the peripheral parts. At $\tau$=1 level, with respect to the dark nuclei this difference is found to be 441 G, 440 G,
900 G and 325 G for UD `a', `b', `c' and the light bridge fragment, respectively. This is in agreement
with findings of Sch\"ussler and V\"ogler (2006). We find higher magnetic field in the peripheral part of these
structures. Joshi {\it et al.} (2007)
reported similar trend in and around UDs in their analytical study on the Joule heating in UDs that
suggests the higher magnetic field in the peripheral part of UDs.

        The line-of-sight velocity stratification for these umbral fine structures is shown in the third column of Figure 4.
UD `a', `b' and `c' show upflow of 300, 280 and 450 ms$^{-1}$, respectively with associated downflow at edges.
The light bridge fragment shows upflow in the middle and downflow at the right edge.

        The fourth column of Figure 4 depicts the inclination stratification for these UDs and the light bridge fragment. We can see
more inclined field above these structures. At $\tau$=1 the field is inclined around 10$^{\circ}$ for UDs `a' and `b' and the light bridge
fragment. However, field is strongly inclined for UD `c', up to 20$^{\circ}$ at $\tau$=1. Such inclined fields forms a cusp above these structures.
Cusp above UDs predicted by Sch\"ussler and V\"ogler (2006) in the simulations. Borrero {\it et al.}
(2008) reported magnetic field wrapping around penumbral filaments. This is
consistent with field geometry we observed for UDs.

      Shown in Figure 5 is the line of sight velocity map of region of interest shown by outer rectangle in Figure 1 for the fast scan at
      log($\tau$)=$-1.5$. The UD `c' shows upward velocity up to 450 ms$^{-1}$. On the other hand other UDs show upflow of the order of
300 ms$^{-1}$which is in agreement with finding of Socas-Navarro
{\it et al.} (2004) and Bharti {\it et al.} (2007a). However we observed down flow patches around larger UD `c'.
 This is in agreement with Bharti {\it et al.} (2007a) who reported downflow around
UDs. However, we haven't observed downflow around smaller UDs. That may be due to the lower spatial resolution
in analysed data for present study. Upward velocity in the central parts and downflow around the peripheral parts of UDs
suggest their magnetoconvective origin as reported by Sch\"ussler and V\"ogler (2006). Since this
sunspot was located close to the disk center hence the line of sight component of
 the velocity is assumed to be vertical.

Shown in  Figure 6 (right) is enlarged Stokes {\it V} area asymmetry map of selected UDs in Figure 3(a). Contours of
continuum intensity image (left) are plotted over area asymmetry map (right). On comparing with rectangular region in velocity
map in Figure 5, one can see that
 upflow region of UD `c' shows positive area asymmetry whereas downflow patches around this UD show negative area
asymmetry. Similar trend can be observed for UDs in a triangular region `d' and the light bridge fragment.
UD `b' shows very interesting pattern of area asymmetry, as shown in Figure 2. It shows bright ring with dark lanes
 around a bright UD. We can see this ring as positive area asymmetry signature with lesser positive
area asymmetry in the center, producing a donut-shaped structure. Auer and Heasley (1978) suggested that to produce an area asymmetry
$\delta${\it A}, a gradient in the velocity is required.  On the other hand, a
combination of gradients in the field strength or orientation and velocity can produce an asymmetry in much more
efficient manner. We observed positive area asymmetry in the upflow region and negative area asymmetry at the downflow
region of UDs, however only brighter UDs show up positive area asymmetry above background noise.

\begin{figure}
\vspace{-5mm}
\centering
\includegraphics[width=90mm,angle=0]{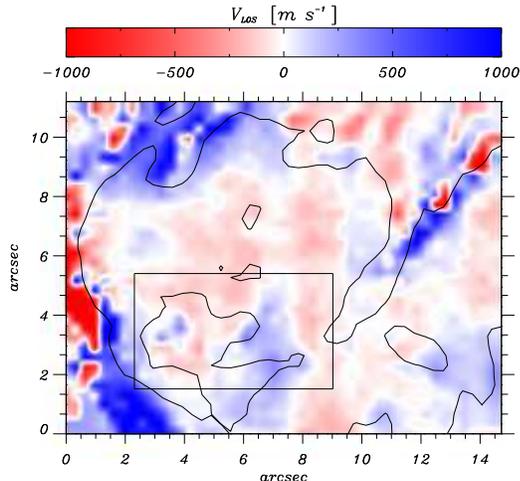}
\vspace{-16mm}
\caption{LOS velocity map derived from inversion for fast scan at log($\tau$)=$-1.5$. UDs show upflow. Bigger UD
`C' as shown in Figure 1 show upflow in central part and downflow patches around it.
Velocity scaled between $\pm$1000 ms$^{-1}$. Blue show upflow and red show downflow.
Contour of continuum intensity is overplotted. Rectangle show location of region of interest subjected to
area asymmetry measurement.}
\end{figure}

\begin{figure}
\vspace{-56mm}
\hspace{11mm}
\includegraphics[width=91mm,angle=0]{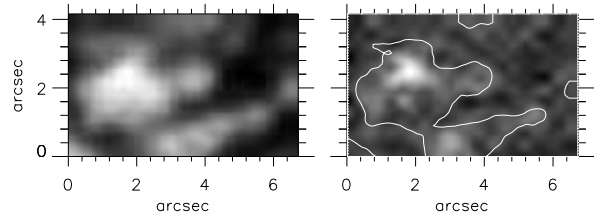}
\vspace{4mm}
\caption{Enlarged view of continuum image of UDs and light bridge fragments (left). Enlarges
view of area asymmetry map of UDs and light bridge fragments (right), contour of
continuum intensity from left image is overplotted. UDs and light bridge fragments clearly
show positive area asymmetry in central part of these structures.}

\end{figure}

\section{Discussion and conclusions}

In this study we present a spectropolarimeteric analysis using sophisticated inversion technique of umbral fine structure that shows dark
lanes in G-band images. Seeing free data from {\it Hinode} with high spatial resolution
of at 0$\prime$$\prime$.6
resolution enables us to study stratification of various parameters associated with umbral fine structure.

Our results support several aspects of 3D MHD simulations with gray
radiative transfer by Sch\"ussler and V\"ogler (2006) such as dark lanes in umbral dots
and the light bridge fragments, temperature deficit in dark lanes, higher LOS magnetic field at
the periphery of UDs, magnetic field reduction in the central part of UDs, more inclined
magnetic field around UDs, upflow in the central part of UDs and downflow in patches around larger UDs.
These facts suggest that UDs appear as a result of magnetoconvection in strong background magnetic field in the sunspot umbra.

Sch\"ussler and V\"ogler (2006) show that UDs have elongated shapes and that there is downflow associated
at the end points of dark lanes. The spatial resolution of our observations is insufficient to confirm
such flow pattern. We observed downflow patches around a larger UD, and a comparison with the dark lanes
in the G-band image, suggests that downflows are not associated with dark lane end points. This larger UD
may have different origin, as a result of flux separation (Weiss {\it et al.} 2002).
Thus, granular-like convection similar to quiet regions appears in a large field-free patches. Hence we could observe
downflows surrounded  around overturning convective cells. However, the stratification of plasma
parameters that we found for UD `c', which displays several dark lanes, is compatible with results of Sch\"ussler and V\"ogler (2006).

 Riethm\"uller et al. (2008a) presented an analysis similar to ours of a large number of central and peripheral
 UDs. They used Hinode normal SP scan at a resolution of (0$\prime$$\prime$.3) and obtained results similar to ours.
 At the log($\tau$)=0 level they found that peripheral UDs on average exhibit a temperature enhancement of 570 K,
a weaker magnetic field of 510 G, and upflow of 800 ms$^{-1}$. On the other
hand, their central UDs on average display a 550 K higher temperature, a weaker field
of 480 G, and no significant upflow signature. However, we find central UD `b' and `c' are hotter by 810 K and 1221 K respectively,
show upflow of a few times 100 ms$^{-1}$ and have a
more inclined field (10$^{\circ}$ and 20$^{\circ}$, respectively). The downflows around the UDs reported
by Riethmuller et al. (2008) are in good agreement with our present study and
confirm the findings of Bharti {\it et al.} (2007a).

The asymmetry in the area that we studied suggests that there are gradients in  the magnetic field, the upflow and downflow velocities, and in the inclination
of the magnetic field. S\'anchez Almida and Lites (1992) suggested the so-called $\triangle$$\gamma$
mechanism, i.e. the simultaneous variation of the velocity and magnetic field inclination to explain area
 asymmetry. Solanki and Montavon (1993) showed sign dependence of an area asymmetry on
 combinations of gradients of these quantities. Equation (2) and (3) of Solanki and Montavon (1993) show the sign of the observed
area asymmetry $\delta$A in and around UDs. These equations are given by:

 \begin{equation}
 sign(\delta A)= -sign(\frac{dV_{los}}{d\tau}. \frac{d|B|}{d\tau}),
 \end{equation}

  \begin{equation}
 sign(\delta A)= -sign(\frac{dV_{los}}{d\tau}. \frac{d|cos\gamma|}{d\tau}).
 \end{equation}

Where V$_{los}$ $>$ 0 for a velocity directed away from the observer.

The central part of UDs show:
{d$\mid$B$\mid$}/{d$\tau$} $<$ 0, {dV$_{los}$}/{d$\tau$} $>$ 0, and {d$\mid$cos$\gamma$$\mid$}/{d$\tau$} $<$ 0,
thus implying positive area asymmetry.
In the peripheral part we find:
{d$\mid$B$\mid$}/{d$\tau$} $>$ 0, {dV$_{los}$}/{d$\tau$} $>$ 0, and {d$\mid$cos$\gamma$$\mid$}/{d$\tau$} $>$ 0
which implies negative area asymmetry.
 Thus, upflow regions show positive area asymmetry and downflow ones show
negative area asymmetry. The gradient we found for these quantities and for the area asymmetry in and around UDs
 are compatible with the model suggested in Figure 2 of Sch\"ussler and V\"ogler (2006). However, due to the
 limit set by the spatial
resolution of our observations we can not observe such asymmetries around smaller UDs (i.e. narrow downflow channels
concentrated at the end points of dark lanes). Sch\"ussler and V\"ogler (2006) also suggested that line
forming region above UDs lies in higher height, thus strong field reduction and high upflow
velocities may not be observable in spectroscopic observation, this is in agreement with our findings.

        Spectropolarimetry of UDs with the Narrowband Filter Imager (NFI) and Dopplergrams
of the magneticaly insensitive line 5576~\AA~ at a 0$\prime$$\prime$.2 will be very useful for
detailed studies of umbral fine structure. On the other hand, observations of umbral
fine structure at different heights in solar atmosphere from ground based facilities will be our next aim.

\section*{acknowledgements}
Juan Manuel Borrero, Jan Jur\v c\'ak and Luis Bellot Rubio (who kindly provided SIR code) are gratefully acknowledged for discussion on SIR inversion.
We thank Prof. Manfred Sch\"ussler and Dr. Michal Sobotka for useful discussions. We indebted to an anonymous referee for useful suggestions to
improve the presentation of this work. Dr. Nick Hoekzema and Dr. Ajay Manglik are gratefully acknowledged for help with language. Hinode is a Japanese mission developed and launched by ISAS/JAXA,
with NAOJ as domestic partner and NASA and STFC (UK) as international partners.
It is operated by these agencies in co-operation with ESA and NSC (Norway).
This research is supported by Bal Shiksha Sadan Samiti (a nongovernmental organization [NGO]
, Udaipur.)

\bsp
\label {lastpage}
\end{document}